\documentclass[12pt]{article}
\usepackage{amsmath,amssymb,epsfig}

\makeatletter

\@addtoreset{equation}{section}
\makeatother

\begin{document}

\title{\vbox{%
\baselineskip 14pt
\hfill \hbox{\normalsize KUNS-2408}\\
} \vskip 1.7cm
\Large \bf Runaway, D term and R-symmetry Breaking
\vskip 0.5cm
}
\author{%
Tatsuo~Azeyanagi,$^1$ \
Tatsuo~Kobayashi,$^2$\\ 
Atsushi~Ogasahara$^2$\,\,and 
Koichi~Yoshioka$^3$
\\*[20pt]
$^1${\it \normalsize
Center for the Fundamental Laws of Nature,  Harvard University, }\\
{\it \normalsize Cambridge Massachusetts 02138, USA} \\
$^2${\it \normalsize
Department of Physics, Kyoto University, Kyoto 606-8502, Japan} \\
$^3${\it \normalsize 
Department of Physics, Keio University, Kanagawa 223-8522, Japan}
}

\date{}

\maketitle
\thispagestyle{empty}
\begin{abstract}
We study the D-term effect on runaway directions of the F-term scalar
potential. A minimal renormalizable model is presented where 
supersymmetry is broken without any pseudomoduli. 
The model is applied to the hidden sector of gauge mediation 
for spontaneously breaking R symmetry and generating nonvanishing gaugino
masses at the one-loop order.
\end{abstract}

\newpage
\baselineskip 17pt
\setcounter{footnote}{0}

\section{Introduction}

Supersymmetry is expected to be one of the key ingredients to describe
physics beyond the Standard Model (SM). While tree-level supersymmetry
breaking within the SM sector leads to light sfermions, the breaking
sector is separated from the SM one and is mediated by some effective
operators or quantum effects. Among various mechanisms for realizing
this scenario, the gauge mediation, relevant to this paper, is one of
the most promising candidates with a strong prophetic power (for a
review, see~\cite{Giudice:1998bp}).

It is known that pseudomoduli directions are present in the
supersymmetry-breaking vacuum of O'Raifeartaigh-like models with the
canonical K\"ahler potential~\cite{Ray:2006wk}. An important implication of this result is that, 
if such models are used as the hidden sector of gauge mediation, 
gaugino masses are generally suppressed or the vacuum is unstable somewhere along the
pseudomoduli~\cite{Komargodski:2009jf}. There have been various ways
in the literature to avoid such a phenomenologically unfavorable
situation, such as including nonminimal terms in the
potential~\cite{Aldrovandi:2008sc}, quantum effects from specific
scalar and/or vector multiplets~\cite{Dine:2006xt}, or accepting
metastable vacua~\cite{Kitano:2006xg}. Another way, as we
discussed before, is to introduce gauge multiplets and take
non-negligible D term into account for making the vacuum stable
without pseudomoduli.

In our previous paper, we classified supersymmetry-breaking models
with nonvanishing F and D terms~\cite{Azeyanagi:2011uc}. First, the
models are divided into two categories based on whether the F-term
potential has a supersymmetric minimum (at finite field
configuration). We then add the D term by gauging flavor symmetry and
analyze the vacuum of the full scalar potential in each category. For
models that do not satisfy the F-flatness conditions, we found that
the full potential generally shows runaway behavior. On the other
hand, when the F-flatness conditions are satisfied, supersymmetry can
be broken without pseudomoduli only in the presence of the 
Fayet-Iliopoulos (FI) term. By using the latter class of models, we
constructed a model of gauge mediation where gaugino masses are
generated at the one-loop order.

In this paper, we discuss another possibility for the classification:
the F-term potential is minimized at some infinite field
configuration, i.e., it shows a runaway behavior. It is found that the
runaway direction of the F-term potential can be uplifted by the D term,
and a supersymmetry-breaking vacuum emerges at finite field
configuration. There are several reasons to explore this class of
models in detail. First of all, contrary to our previous result,
there is no need to add the FI term for supersymmetry
breaking.\footnote{The FI term also has some difficulty when 
incorporated into local supersymmetric theory~\cite{Komargodski:2009pc}.}
Secondly, the vacuum automatically suppresses pseudomoduli directions
associated with the F-term potential, since it has a runaway behavior
and is stabilized by the D-term potential. We propose a minimal model 
with such properties and couple it with an appropriate messenger sector. 
In this model including the messenger sector, R symmetry is spontaneously broken 
at the tree level even though the model contains chiral superfields with U(1)$_R$ charge 0 or 2 only. 
We notice that R symmetry breaking does not occur for O'Raifeartaigh-like models with 
such a U(1)$_R$ charge assignment \cite{Shih:2007av}.
This class of supersymmetry breaking can therefore provide a realistic model for gauge mediation, 
where leading-order gaugino masses are obtained at the stable vacuum.

The outline of this paper is as follows. In Sec.~\ref{sec:review},
we briefly review our classification of F- and D-term supersymmetry
breaking. In Sec.~\ref{sec:runaway}, we discuss the case in which the
F-term potential shows runaway behaviors. After some general
arguments, a minimal model is presented to realize the vacuum property
listed above. Further, appropriate messenger sectors are discussed and
shown to be viable for generating gaugino
masses. Section~\ref{sec:conclusion} is devoted to summarizing our
results and discussions on future directions. In the Appendix, we show
the potential analysis of the model given in Sec.~\ref{sec:runaway}.
\medskip

\section{Supersymmetry breaking with F and D terms}
\label{sec:review}

We first review our previous result of the classification of
supersymmetry breaking with both F and (Abelian) D
terms~\cite{Azeyanagi:2011uc}. Throughout this paper, we assume the
K\"ahler potential is canonical. The superpotential $W$ has a
polynomial form of chiral superfields $\phi_i$ with U(1) charges $q_i$ (the
latin indices label their species). The scalar potential $V$ is then given by
\begin{eqnarray}
  V \,=\, V_F + V_D \,,
\end{eqnarray}
where $V_F$ and $V_D$ are the contributions from F and D terms:
\begin{alignat}{2}
  & V_F \,=\, \sum_i |F_i|^2, 
  &\qquad& F_i \,=\, 
  -\left(\frac{\partial W}{\partial \phi_i}\right)^*, \\
  & V_D \,=\, \frac{g^2}{2}D^2,
  &\qquad& D \,=\, \sum_i q_i|\phi_i|^2+\xi \,.
\end{alignat}
Here $g$ is the U(1) gauge coupling constant and $\xi$ is the
coefficient of the possible Fayet-Iliopoulos term~\cite{Fayet:1974jb}. In
the following, we abbreviate field derivatives of the superpotential 
as $W_{\phi_i}$ ($=\partial W/\partial\phi_i$).

We first divide models into two categories. The criterion for the
classification is whether the F-flatness condition, $V_F=0 $, is
satisfied or not at its minimum defined 
by $\partial V_F/\partial\phi_i=0$. If the F-flatness condition is
(not) satisfied, a model is called in the second (first) class. We
then add the U(1) D term and analyze the tree-level behavior of the
full scalar potential. For the first class of models, the scalar
potential shows supersymmetric runaway behaviors when the D-term
contribution is included.\footnote{A similar behavior was studied in
an explicit example in \cite{Matos:2009xv}} For the second class of
models, on the other hand, supersymmetry can be broken without any
pseudomoduli. That is, however, realized only in the presence of the FI
term. (See Table~\ref{table:class} for the classification.)

\begin{table}[t]
\begin{center}
\begin{tabular}{l|ccc} \hline\hline
  & ~ At $\frac{\partial V_F}{\partial \phi_i}=0$ ~ & 
FI term ($\xi$) $=0$  & ~ FI term ($\xi$) $\neq0$ ~ \\ \hline
First class  & $V_F\not= 0$ & runaway & runaway \\[1mm]
Second class ~ & $V_F=0$ & SUSY & SUSY breaking \\ \hline
\end{tabular}\medskip
\caption{Classification of supersymmetry breaking with F and D terms.}
\label{table:class}
\end{center}
\end{table}

There is another possibility in the classification, which we did not
consider previously: the F-term potential $V_F$ has runaway behavior,
i.e., the minimization $\partial V_F/\partial\phi_i=0$ is not
satisfied (with finite field configuration). In the following, we
focus on this class of models, giving some general arguments and
discussing a minimal model that shows that the runaway direction is
uplifted by the D-term potential $V_D$. The model also has the property 
that pseudomoduli are absent in the vacuum. 
By coupling it with an appropriate messenger sector, 
we show that R symmetry is spontaneously broken and gaugino masses are
generated at the one-loop order.

\medskip

\section{Runaway and D-term uplift }
\label{sec:runaway}

\subsection{General arguments}

A well-known runaway behavior of the scalar potential arises from a
nonperturbative superpotential which has inverse powers of field
variables. Supersymmetry breaking occurs along such a runaway direction
if a suitable superpotential is added to lift it up~\cite{Affleck:1983mk}. It is, however, noted in this case
that the D-term potential is vanishing.\footnote{Deviation from 
D-flat directions has been discussed for several models
recently~\cite{Elvang:2009gk}.} On the other hand, there is another
type of runaway behavior related to symmetries of theory. We are
interested in how the runaway potential is affected by a
nonvanishing D term, which is given by gauging non-R flavor
symmetry. In this paper, we focus on the Abelian D term, but
non-Abelian generalization is straightforward.
 
We first see how the runaway directions related to U(1) gauge symmetry
occur. Consider a theory with U(1) (gauge) symmetry. The
superpotential is then invariant under the complexified U(1) which
contains a charge-dependent scale transformation as its real part. An
important point is that the F-flatness conditions are satisfied with
the variables obtained by the U(1) transformation from the original
solution. Let us assume that there exists a field 
configuration $\phi_i=\phi_i^{(0)}$ that realizes the following
situation: $F_i=0$ for all fields with seminegative 
charges ($q_i\le 0$),\footnote{One can relax the condition such 
that $F_i\neq 0$ for some neutral fields. The following arguments of
runaway behavior still hold, since neutral fields and their F terms
are unchanged along the runaway direction. The only difference is the
size of the F-term potential at infinity of the field space.}
while $F_j\neq 0$ for at least one field with positive 
charge ($q_j>0$). Under the above scale transformation acting 
on $\phi_i^{(0)}$,
\begin{equation}
  \phi_i^{(0)} \to\, e^{q_i \alpha}\phi_i^{(0)} 
  \qquad (\alpha\in {\mathbb R}) \,, \label{eq:runaway}
\end{equation}
the F terms behave as
\begin{alignat}{2}
  W_{\phi_i} &\,=\; 0 & \qquad (q_i\leq0) \,, \\
  W_{\phi_i} &\,\to\; e^{-q_i\alpha}W_{\phi_i} &\qquad (q_i>0) \,,
\end{alignat}
and the F-term potential approaches to zero as $\alpha\to\infty$. At the
same time, some values of positively charged fields go to
infinity. The F-term potential satisfying the above assumption, therefore, shows a runaway behavior along the direction related to the U(1)
symmetry. Similarly, the runaway also occurs if scalar fields realize
the following situation: $F_i=0$ for all fields with semipositive
charges ($q_i\geq0$), while $F_j\neq 0$ for at least one field with
negative charge ($q_j<0$). In this case, the runaway direction
corresponds to $\alpha\to-\infty$.\footnote{The runaway behavior
related to U(1)$_R$ symmetry is understood in a similar
way~\cite{Ferretti:2007ec}. A difference from non-R symmetry discussed
in this section is that the superpotential carries a nonzero charge,
and one can specify R-charge assignment for runaway to occur.}

Then we include the U(1) D term and examine how the scalar potential
is modified. If the charge-dependent scale transformation is applied
and the parameter $\alpha$ is taken to infinity, the D term behaves as
\begin{equation}
  D \,\to\, \sum_{q_i>0}\,q_i|\phi_i|^2 \qquad (\alpha\to\infty) \,.
\end{equation}
Similarly, for $\alpha\to-\infty$, the D term is dominated by
negatively charged fields. Along the U(1) runaway direction discussed
above, the D term grows up as $|\alpha|\to\infty$ and uplifts the full
scalar potential away from the origin. This result is irrelevant to
whether a nonvanishing FI term exists or not. We here comment on an
important point that since a runaway F-term potential does not satisfy
its stationary conditions at finite field configuration, the theory is
expected to have no pseudomoduli by formulation.

It is noted that while the U(1) runaway direction is stabilized, the
full potential is not necessarily stabilized in the direction~\eqref{eq:runaway}. For
example, one may have other directions, such as U(1)$_R\,$-related
runaway, orthogonal to the stabilized direction. Moreover, the
potential must be carefully analyzed so that it is indeed minimized,
not a saddle point. A more general argument seems hard to complete,
and we leave these issues to future work. In the following sections, 
we present a minimal tree-level model with F-term runaway lifted by the
D term of gauged flavor symmetry.

\medskip

\subsection{A minimal model}
\label{subsec:example}

In this section we study a renormalizable model satisfying the
assumption we have made above: symmetry-related runaway directions of
the F-term potential can be uplifted by the D term, and a
supersymmetry-breaking vacuum emerges at finite field
configuration. The model is explicitly shown to have a stable vacuum
with no pseudomoduli, where supersymmetry is broken. By coupling it
with an appropriate messenger sector, we show that R symmetry is spontaneously broken and gaugino masses are
generated at the one-loop order.

\subsubsection{Supersymmetry-breaking sector and runaway}

The model consists of six chiral 
multiplets $X_\pm$, $X_0$, $\varphi_\pm$ and $\varphi_0$, where the
subscripts denote their U(1) charges. The assignment 
of U(1) and U(1)$_R$ charges is summarized below.
\begin{table}[h]
\begin{center}
\begin{tabular}{c|cccccc} \hline\hline
& $X_+$ & $X_-$ & $X_0$ & $\varphi_+$ & $\varphi_-$ & $\varphi_0$ \\ \hline
U(1) & $+1$ & $-1$ & 0 & $+1$ & $-1$ & 0 \\
U(1)$_R $ & 2 & 2 & 2 & 0 & 0 & 0 \\ \hline
\end{tabular}\medskip
\caption{The assignment of U(1) and U(1)$_R$ charges.}
\label{table:assign}
\end{center}
\end{table}\\
The superpotential is 
\begin{equation}
  W \,=\, f X_0+\lambda\,X_0\,\varphi_+\varphi_- 
  +m\,X_+\varphi_- +\lambda'\,X_-\varphi_+\varphi_0 \,.
  \label{example_superpotential}
\end{equation}
This form can be the most generic, renormalizable one
if $\varphi_0$ and $X_-$ have the charges $+1$ and $-1$ under an 
additional $Z_N$ symmetry ($N>2$). We here comment on the role of each
term in the superpotential~\eqref{example_superpotential}: the first
and second terms make the origin unstable along the 
meson $\varphi_+\varphi_-$. The third term lifts up 
the $\varphi_-$ direction, and naively supersymmetry is broken with
the F-term potential coming from the first three terms. However, the
potential minimum is found to run away to infinity of the moduli space
along the $\varphi_+$ direction (with a finite value 
of $\varphi_+\varphi_-$). It is further noticed that anomaly
cancellation requires the existence of a negatively charged 
field ($X_-$). Without the last fourth term, the value of $X_-$ is
free and the direction $X_-\varphi_+$ becomes D flat, along which the
potential minimum goes to infinity and supersymmetry is recovered. In
this way, the superpotential \eqref{example_superpotential} is
regarded as minimal one for the present purpose.

The scalar potential $V=V_F+V_D$ is explicitly given by 
\begin{eqnarray}
  && V_F \,=\, |f+\lambda\varphi_+\varphi_-|^2 +|m\varphi_-|^2
  +|\lambda'\varphi_+\varphi_0|^2  \nonumber \\
  && \qquad\qquad +|\lambda'X_-\varphi_+|^2
  +|\lambda X_0\varphi_-+\lambda'X_-\varphi_0 |^2
  +|\lambda X_0\varphi_++mX_+|^2 \,,
  \label{fterm_pot}  \\[1mm]
  && V_D \,=\, \frac{g^2}{2}D^2 \,=\,
  \frac{g^2}{2}\big(|X_+|^2+|\varphi_+|^2-|X_-|^2-|\varphi_-|^2\big)^2 \,.
  \label{dterm_pot}
\end{eqnarray}
The F-term potential $V_F$ has the U(1) runaway directions discussed
in the previous section. We find along the following direction
\begin{eqnarray}
  X_+=X_-=X_0=\varphi_0=0 \,, \qquad
  \varphi_+=\sqrt{\frac{f}{\lambda}}\,e^\alpha , \qquad 
  \varphi_-=-\sqrt{\frac{f}{\lambda}}\,e^{-\alpha},
  \label{runaway_config}
\end{eqnarray}
and all the F terms vanish except for the positively charged 
field $F_{X_+}$, 
\begin{eqnarray}
  F_{X_-}=F_{X_0}=F_{\varphi_+}=F_{\varphi_-}=F_{\varphi_0}=0 \,, \qquad
  F_{X_+}^*=m\sqrt{\frac{f}{\lambda}}\,e^{-\alpha} \,.
\end{eqnarray}
The runaway direction is parametrized 
by $\alpha$. As $\alpha\to\infty$, the positively charged 
field $\varphi_+$ goes to infinity and the F-term 
potential ($F_{X_+}$) approaches to zero.  Furthermore, no U(1)$_R$ runaway is expected
since all the scalar fields in the present model have R charges 0 or
2~\cite{Ferretti:2007ec}.

\begin{figure}[t]
\begin{center}
\includegraphics[width=6.5cm]{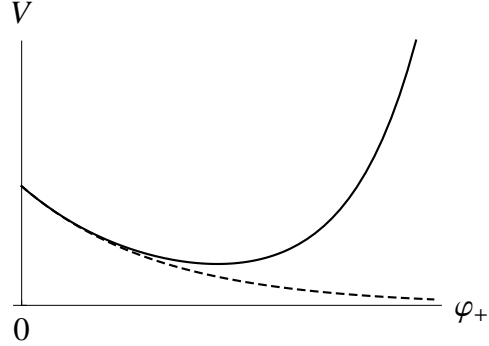}\medskip
\caption{A typical behavior of $V_F$ (dashed line) and $V$ (solid line)
along the runaway direction \eqref{runaway_config}. The runaway
direction of $V_F$ is uplifted by the D-term contribution $V_D$.}
\label{fig:uplift}
\end{center}
\end{figure}

It is easily seen that the F-term runaway 
direction \eqref{runaway_config} is stabilized by the D-term
contribution because, along this direction, the D term increases as
\begin{equation}
  D \,\to\, \big|\varphi_+\big|^2 \qquad (\alpha\to\infty) \,.
\end{equation}
(See also Fig.~\ref{fig:uplift}.) \ We again note that even when the
runaway is lifted by the D term, it does not necessarily mean that
the vacuum of the scalar potential is in the 
direction~\eqref{runaway_config}.

\medskip

\subsubsection{Supersymmetry-breaking vacuum}

We then analyze the scalar potential in detail to confirm that a
stable supersymmetry-breaking vacuum is obtained in some parameter
region. The parameters appearing in the superpotential are assumed to
be real and positive without loss of generality.

The vacuum is identified by solving the stationary conditions for the
full scalar potential $V$. A trivial configuration satisfying the
stationary conditions is the origin at which all the scalar fields
vanish. This point however is unstable and we do not consider it in
the following. By some calculation (the detail is summarized in
the appendix), we can show that the vacuum satisfies
\begin{equation}
  X_+ =\, X_0 \,=\, \varphi_0 \,=\, 0 \,,
  \label{vacuum0}
\end{equation} 
where several F terms vanish; 
\begin{equation}
  F_{X_-} \,=\, F_{\varphi_+} \,=\, F_{\varphi_-} \,=\, 0 \,.
  \label{Fterms}
\end{equation}
Then the stationary conditions 
for $X_+$, $X_0$, and $\varphi_0$ automatically hold, and the scalar
potential simplifies to
\begin{equation}
  V \,=\, |f+\lambda\varphi_+\varphi_-|^2 +|m\varphi_-|^2
  +|\lambda'X_-\varphi_+|^2 +\frac{g^2}{2}D^2 \,,
\end{equation}
with $D=|\varphi_+|^2-|\varphi_-|^2-|X_-|^2$. ~With this reduced
potential, the stationary condition for $X_-$ reads
\begin{eqnarray}
  \frac{\partial V}{\partial X_-^*} \,=\, 
  X_-\big(\,|\lambda'\varphi_+|^2-g^2D\big) \,=\, 0 \,, 
\end{eqnarray}
indicating $X_-=0\,$ or $\,D=|\lambda'\varphi_+/g|^2$. We discuss
these two cases separately below, solving the remaining stationary
conditions for $\varphi_\pm$.

\medskip

\underline{(1) $X_-=0$ } : In this case, for a large 
mass ($m^2\gg\lambda f,\, \lambda^2f/g$) or a small 
one ($m^2\ll\lambda f,\, g^2f/\lambda$), we can approximately write
down the analytic solution to the stationary conditions 
for $\varphi_\pm$. For the large mass regime, the solution is given by
\begin{equation}
  \qquad\quad
  \varphi_+ \,=\, \pm\frac{\lambda f}{gm} \,, \qquad 
  \varphi_- \,=\, \mp\frac{\lambda^2f^2}{gm^3} \,,
  \qquad (\text{large }\, m)
\end{equation}
where the F and D components other than \eqref{Fterms} become
\begin{equation}
  F_{X_+} =\, \pm\frac{\lambda^2f^2}{gm^2} \,, \qquad 
   F_{X_0} \,=\, -f+\frac{\lambda^4 f^3}{g^2m^4}\,, \qquad 
  F_{\varphi_0} \,=\, 0 \,, \qquad
  D \,=\, \frac{\lambda^2f^2}{g^2m^2} \,,
  \label{X-=0_large_m}
\end{equation}
at the leading order. Therefore, the scalar potential is dominated 
by $F_{X_0}$, i.e., $V\simeq f^2$. On the other hand, for the small
mass regime, we have
\begin{eqnarray}
  \qquad\qquad
  \varphi_+ \,&=&\, \pm\sqrt{\frac{f}{\lambda}}\left(1-\frac{m^2}{4\lambda f} +\frac{\lambda m^2}{8g^2f}\right) \,, \\
  \varphi_- \,&=&\, \mp\sqrt{\frac{f}{\lambda}}\left(1-\frac{m^2}{4\lambda f} -\frac{\lambda m^2}{8g^2f}\right) \,,
  \qquad (\text{small }\, m)
 \end{eqnarray}
where the F and D components are
\begin{eqnarray}
  F_{X_+} =\, \mp m\sqrt{\frac{f}{\lambda}} \,, \qquad 
  F_{X_0} \,=\, -\frac{m^2}{2\lambda} \,, \qquad 
  F_{\varphi_0} \,=\, 0 \,, \qquad
  D \,=\, \frac{m^2}{2g^2} \,,
  \label{X-=0_small_m}
\end{eqnarray}
and the scalar potential is found to be dominated 
by $F_{X_+}$, i.e., $V\simeq fm^2/\lambda$. Notice that, in both
regimes of $m$, the R symmetry is unbroken as $F_{\varphi_0}=0$.

\begin{figure}[t]
\begin{center}
\includegraphics[width=6.5cm]{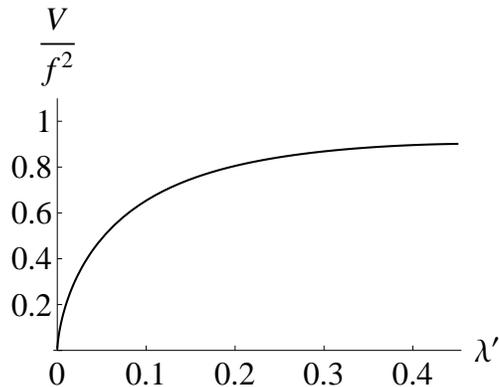}\medskip
\caption{The potential value at the vacuum as a function 
of $\lambda'$ (for the $D=|\lambda'\varphi_+/g|^2$ case). The other
parameters are fixed to $(m^2/f,\,\lambda,\,g)=(2,\,0.7,\,0.5)$. In
the limit $\lambda'\to0$, supersymmetry is recovered, but some
expectation values run away to infinity.}
\label{fig:vac-height}
\end{center}
\end{figure}

The stability of these vacua is read off from the eigenvalues of the
squared mass matrix for the scalar fields. We find that all eigenvalues
except for $X_-$ are positive semidefinite. Along the $X_-$ direction, the
eigenvalue is given by
\begin{equation}
  M^2_{X_-} \,=\, (\lambda'^{\,2}-g^2)|\varphi_+|^2 +g^2|\varphi_-|^2 \,.
\end{equation}
This eigenvalue is positive when $\lambda'\gtrsim g$ for the large
mass regime and $\lambda'\gtrsim (\lambda m^2/f)^{1/2}$ for the small
mass regime. Therefore the supersymmetry-breaking vacuum realized is
stable for these parameters. Otherwise, $X_-=0$ is a saddle point and
the true vacuum is given by the following second case:

\medskip

\underline{(2) $D=|\lambda'\varphi_+/g|^2$ } : In this case, the
stationary conditions are complicated and we only present a typical
numerical solution. For example, when the model parameters are set 
as $(m^2/f,\,\lambda,\,\lambda',\,g)=(2,\,0.7,\,0.1,\,0.5)$, which
do not satisfy the above condition for the stability of $X_-=0$, the
vacuum is located at
\begin{equation}
  (\,\varphi_+,\,\varphi_-,\, X_-) \,\simeq\, 
  (1.34,\,-0.252,\,1.29)\times m \,,
\end{equation}
where nonvanishing F and D components are
\begin{equation}
  \big(F_{X_+},\,F_{X_0},\,F_{\varphi_0},\,D\big) \,\simeq\, 
  (-0.504,\, -0.527,\, 0.346,\, 0.144)\times f \,.
\end{equation}
All the F and D terms become comparable to each other and contribute
to the scalar potential $V\sim{\cal O}(f^2)$. The stability of this
vacuum is confirmed numerically. We have also checked that, for a wider
parameter region, the scalar potential has a similar behavior. In 
Fig.~\ref{fig:vac-height}, we show the normalized 
potential $V/f^2$ characterizing supersymmetry breaking at the vacuum
as a function of $\lambda'$ for the large mass 
regime ($m^2\gg\lambda f,\, \lambda^2f/g)$. Without 
the $\lambda'$ term, supersymmetry recovers at infinity of the moduli
space, as we mentioned before.
As $\lambda'$ becomes larger, $X_-$ is stabilized 
for $\lambda'\gtrsim g$ and the vacuum is shifted up 
to \eqref{X-=0_large_m}, where $V/f^2\simeq1$.

\medskip

\subsubsection{Messenger sector and gaugino masses}

We then discuss the gauge-mediation scenario by employing the above
model as a supersymmetry breaking sector. Appropriate messenger fields
and their superpotential are identified for generating one-loop-order
gaugino masses for the two 
cases, $X_-=0$ and $D=|\lambda'\varphi_+/g|^2$, separately.

\medskip

\underline{(1) $X_-=0$ } : As we have shown, the F and D terms have
different behaviors depending of whether the mass parameter $m$ is large or
small. For the large $m$ case, \eqref{X-=0_large_m}, the
supersymmetry-breaking scale is governed by $F_{X_0}$ from which
sfermions are expected to receive soft masses. For generating a
similar size of gaugino masses, a simple way is to introduce the
messenger fields $M$ and $\tilde{M}$ which are vectorlike under the
SM gauge symmetry and have the superpotential,
\begin{equation}
  W \,=\, X_0M\tilde{M} +m_MM\tilde{M}\,.
\end{equation}
For $ f\ll m_M^2$,  the standard one-loop diagram of messenger fields generates a gaugino
mass $M_g$;
\begin{equation}
  M_g \,=\, \frac{T_R g_\text{SM}^2}{8\pi^2}\,\frac{f}{m_M} \,,
  \label{gauginomass1}
\end{equation}
where $g_\text{SM}$ is the SM gauge coupling, and $T_R$ is the Dynkin
index of the SM gauge symmetry for $M$ and $\tilde{M}$. The gaugino
mass \eqref{gauginomass1} is given at the vacuum discussed in the
previous section. It is, however, noticed that the R symmetry is softly
broken by the parameter $m_M$ that makes $X_0=0$ a local minimum. Another way of obtaining
gaugino masses would be to consider the direct gauge mediation, i.e.,
to generalize U(1) to non-Abelian symmetry containing the SM
one. By adding a small supersymmetric mass for $\varphi_\pm$, they behave
as messengers and would induce the SM gaugino masses without
introducing additional multiplets.

For the small mass case, \eqref{X-=0_small_m}, the charged F 
term $F_{X_+}$ dominates the supersymmetry-breaking scale. To split 
messenger masses with $F_{X_+}$, we must introduce two pairs of
vectorlike messengers with U(1) charges $\pm q$ and $\pm(q-1)$ and
couple them with $X_+$ in the superpotential. For further details of
the messenger sector, the stability of the vacuum, and gaugino mass 
generation, see Ref.~\cite{Azeyanagi:2011uc}.

\medskip

\underline{(2) $D=|\lambda'\varphi_+/g|^2$ } : In the vacuum, the U(1)$_R$
symmetry given in Table \ref{table:assign} is broken by the F term of $\varphi_0$ which is neutral under
both U(1) and U(1)$_R$. We consider the following messenger superpotential 
\begin{eqnarray}
  W \,=\, \varphi_0M\tilde{M}+m_MM\tilde{M} \,, \label{messenger}
\end{eqnarray}
where $M$ and $\tilde{M}$ are vectorlike multiplets, charged
under the SM gauge symmetry. It is noticed that the $Z_N$ symmetry is softly broken in 
this messenger sector. 
Note that the superpotential (\ref{example_superpotential}) has an anomalous U(1)' symmetry, 
under which $\varphi_0$ and $X_-$ are charged, but it is explicitly broken in (\ref{messenger}).
The R symmetry is spontaneously broken in the full potential with both (\ref{example_superpotential}) and (\ref{messenger}), 
if the vacuum in the previous section is stable.
The vacuum stability is ensured by taking the parameter $m_M$ sufficiently large.
The gaugino mass is evaluated for
$f\ll m_M^2$ as
\begin{equation}
  M_g \,=\, \frac{T_Rg_{SM}^2}{8\pi^2} \frac{F_{\varphi_0}}{m_M}\, ,
\end{equation}
which comes from a one-loop diagram, where $M$, $\tilde{M}$ circulate
in the loop.

\medskip

From these observations, we conclude that this class of
supersymmetry-breaking models is useful to build a simple, realistic
hidden sector of gauge mediation.

\medskip

\section{Summary and discussions}
\label{sec:conclusion}

We have studied supersymmetry-breaking models with both F and D terms
being nonvanishing. In particular, we have focused on the case that
the F-term potential shows runaway behaviors that originate from
symmetries of theory considered. The runaway directions are uplifted
by the D term and a supersymmetry-breaking vacuum is realized at
finite field configuration. An interesting property of this approach
is that (phenomenologically disfavored) pseudomoduli are absent in
the vacuum since they are related to the minimization of the F-term
potential. Moreover, there is no need to add the FI term for supersymmetry breaking. 
Along this line, a minimal renormalizable model has been
presented where supersymmetry is broken. For an application to gauge mediation, we have
introduced appropriate messenger sectors and confirmed that R symmetry is spontaneously broken, and gaugino
masses in the visible sector are generated at the comparable order of
sfermion masses. This class of models may open up a new way to build
realistic models of gauge mediation, circumventing the lemma proved by
Komargodski and Shih~\cite{Komargodski:2009jf}.

As remarked, the D-term lifted runaway might be destabilized along other
orthogonal directions such as the U(1)$_R$ runaway. We do not have any
criteria to ensure that the D-term uplifting of runaway directions can
lead to the stable and global minimum of the scalar potential. It would be
interesting if one could carry out a general argument on this issue.

\bigskip

\subsection*{Acknowledgements}

We would like to thank L.B.~Anderson for useful advice on the numerical
analysis. T.A.\ thanks the YITP Workshop on String Theory and Field Theory and 
the 2012 Simons Workshop on Mathematics and Physics, where this work 
was partly done, for hospitality.
T.A.\ is in part supported by the JSPS Postdoctoral Fellowship
for Research Abroad and is grateful to the Center for the Fundamental Laws
of Nature at Harvard University for support. K.Y.\ is supported in
part by the Grant-in-Aid for Scientific Research No.~23740187 and also
in part by Keio Gijuku Academic Development Funds. This work is
supported by ``The Next Generation of Physics, Spun from
Universality and Emergence," the GCOE Program from the Ministry of Education, Culture,
Sports, Science and Technology of Japan.

\newpage

\appendix

\section{Potential analysis}
\label{appendix:potential}

In this Appendix, we give some details of the potential analysis
for the model given in Sec.~\ref{subsec:example}. In particular, we
explain the derivation of the vacuum expectation values \eqref{vacuum0}.  

The field derivatives of the superpotential
\eqref{example_superpotential} are given by  
\begin{gather}
  W_{X_+} = m\varphi_- \,, \qquad
  W_{X_-} = \lambda'\varphi_+\varphi_0 \,, \qquad
  W_{X_0} = f+\lambda\varphi_+\varphi_- \,, \quad \nonumber \\[1mm]
  W_{\varphi_+} = \lambda X_0\varphi_- +\lambda'X_-\varphi_0 \,, \qquad
  W_{\varphi_-} = \lambda X_0\varphi_+ +mX_+ \,, \qquad
  W_{\varphi_0} = \lambda'X_-\varphi_+ \,.
\end{gather}
The scalar potential $V$ is the sum of the contributions from F and D
terms, $V_F$ and $V_D$, and these explicit forms are written down
in \eqref{fterm_pot} and \eqref{dterm_pot}.  
The stationary conditions for $V$ are then given by
\begin{align}
  \frac{\partial V}{\partial X_+^*} \,=\;\,
  & m^* (\lambda X_0\varphi_+ +mX_+) +g^2 X_+ D \,=\, 0 \,, \label{x+sta} \\
  \frac{\partial V}{\partial X_-^*} \,=\;\,
  & |\lambda'\varphi_+|^2X_- 
  +\lambda'^*\varphi_0^*(\lambda X_0\varphi_- +\lambda'X_-\varphi_0)
  -g^2X_- D \,=\, 0 \,,  \label{x-sta} \\
  \frac{\partial V}{\partial X_0^*} \,=\;\,
  & \lambda^*\varphi_+^* (\lambda X_0\varphi_+ +mX_+)
  +\lambda^*\varphi_-^* (\lambda X_0\varphi_- +\lambda'X_-\varphi_0) 
  \,=\, 0 \,,  \label{x0sta} \\ 
  \frac{\partial V}{\partial \varphi_+^*} \,=\;\,
  & \lambda^*\varphi_-^* (f+\lambda\varphi_+\varphi_-)
  +|\lambda'X_-|^2\varphi_+  \nonumber \\ 
  & \quad +\lambda^*X_0^*(\lambda X_0\varphi_+ +mX_+) 
  +|\lambda'\varphi_0|^2\varphi_+ +g^2\varphi_+ D \,=\, 0 \,, \\ 
  \frac{\partial V}{\partial \varphi_-^*} \,=\;\,
  & \lambda^*\varphi_+^* (f+\lambda\varphi_+\varphi_-)
  +|m|^2\varphi_-  \nonumber \\ 
  & \quad +\lambda^*X_0^*(\lambda X_0\varphi_- +\lambda'X_-\varphi_0)
  -g^2\varphi_- D \,=\, 0 \,, \\
  \frac{\partial V}{\partial \varphi_0^*} \,=\;\,
  & \lambda'^* X_-^* (\lambda X_0\varphi_- +\lambda'X_-\varphi_0)
  +|\lambda'\varphi_+|^2\varphi_0 \,=\, 0 \,.  \label{ph0sta}
\end{align}
First, by using \eqref{x0sta} and \eqref{ph0sta}, we express $X_0$ and
$\varphi_0$ in terms of the other fields:
\begin{align}
  X_0 & \,=\; \frac{-m}{\lambda}\,
  \frac{X_+\big(|X_-|^2+|\varphi_+|^2\big)}{\varphi_+\big(|X_-|^2
    +|\varphi_+|^2+|\varphi_-|^2\big)} \,,  \label{x0} \\[1mm]
  \varphi_0 & \,=\; \frac{m}{\lambda'}\,
  \frac{X_+X_-^*\varphi_-}{\varphi_+\big(|X_-|^2
    +|\varphi_+|^2+|\varphi_-|^2\big)} \,.  \label{ph02}
\end{align}
If $X_+\neq0$, we find from \eqref{x+sta} and \eqref{x0},
\begin{equation}
  g^2D \,=\, \frac{-|m\varphi_-|^2}{|X_-|^2
    +|\varphi_+|^2+|\varphi_-|^2} \,.  \label{dneg}
\end{equation}
This implies that $D$ is negative at the stationary point (if $X_+\neq0$). 
On the other hand, from \eqref{x-sta} and \eqref{ph0sta}, we have
\begin{equation}
  |\lambda' X_-\varphi_+|^2 \,=\, 
  |\lambda'\varphi_+ \varphi_0|^2 +g^2|X_-|^2 D \,, \label{rewritten3}
\end{equation}
while the equation $\varphi_+^*(\partial V/\partial \varphi_+^*)
-\varphi_-^*(\partial V/\partial \varphi_-^*)=0$ gives 
\begin{eqnarray}
  && |\lambda'X_-\varphi_+|^2+\varphi_+|\lambda'\varphi_+\varphi_0|^2
  -|m\varphi_-|^2 +g^2 (|\varphi_+|^2+|\varphi_-|^2) D  \nonumber \\[1mm]
  && \qquad -\varphi_-\lambda X_0 (\lambda X_0\varphi_- 
  +\lambda'X_-\varphi_0)^* +\lambda X_0(\lambda X_0\varphi_+ +mX_+)^*
  \,=\, 0 \,. \label{ph+ph-}
\end{eqnarray}
By using the relations,
\begin{align}
  \lambda X_0 (\lambda X_0\varphi_+ +mX_+)^* & =\, 
  \frac{-|mX_+\varphi_-|^2}{\varphi_+}
  \frac{|X_-|^2+|\varphi_+|^2}{(|X_-|^2+|\varphi_+|^2+|\varphi_-|^2)^2}
  \, , \\
  \lambda X_0 (\lambda X_0\varphi_- +\lambda'X_-\varphi_0)^* & =\,
  \frac{|mX_+\varphi_-|^2}{\varphi_-}
  \frac{|X_-|^2+|\varphi_+|^2}{(|X_-|^2+|\varphi_+|^2+|\varphi_-|^2)^2} \,,
\end{align}
following from \eqref{x0} and \eqref{ph02}, 
we find another expression for $D$ from \eqref{rewritten3} and \eqref{ph+ph-}:
\begin{equation}
  g^2D \,=\, \frac{1}{|X_-|^2+|\varphi_+|^2+|\varphi_-|^2}
  \bigg[\,|m\varphi_-|^2 +\frac{2|mX_+\varphi_+\varphi_-|^2}{(|X_-|^2
    +|\varphi_+|^2+|\varphi_-|^2)^2}\bigg] \,. \label{dpos}
\end{equation}
That implies $D$ is positive at the stationary point.
For both conditions \eqref{dneg} and \eqref{dpos} to be true,
$D=0$ is the only solution, but it cannot be satisfied because the origin
of meson direction $\varphi_+\varphi_-$ is destabilized by the F-term
potential. Therefore $X_+\neq0$, which is the assumption for
\eqref{dneg}, should not be realized. In the end, from \eqref{x0} and
\eqref{ph02}, we find the following vacuum expectation values \eqref{vacuum0}:
\begin{equation}
  X_+ =\, X_0 \,=\, \varphi_0 \,=\, 0 \,. \nonumber
\end{equation}


\newpage

\begin{thebibliography}{99}

\bibitem{Giudice:1998bp}
  G.~F.~Giudice and R.~Rattazzi,
  Phys.\ Rept.\  {\bf 322} (1999) 419
  [arXiv:hep-ph/9801271].


\bibitem{Ray:2006wk}
  S.~Ray,
  Phys.\ Lett.\ B {\bf 642} (2006) 137
  [arXiv:hep-th/0607172].


\bibitem{Komargodski:2009jf}
  Z.~Komargodski and D.~Shih,
  JHEP {\bf 0904} (2009) 093
  [arXiv:0902.0030 [hep-th]].


\bibitem{Aldrovandi:2008sc}
  L.~G.~Aldrovandi and D.~Marques,
  JHEP {\bf 0805} (2008) 022
  [arXiv:0803.4163 [hep-th]];
%
  Y.~Nakai and Y.~Ookouchi,
  JHEP {\bf 1101} (2011) 093
  [arXiv:1010.5540 [hep-th]];
%
  T.~S.~Ray,
  Phys.\ Rev.\ D {\bf 85} (2012) 035003
  [arXiv:1111.4266 [hep-ph]].


\bibitem{Dine:2006xt}
  M.~Dine and J.~Mason,
  Phys.\ Rev.\ D {\bf 77} (2008) 016005
  [hep-ph/0611312];
%
  K.~Intriligator and M.~Sudano,
  JHEP {\bf 1006} (2010) 047
  [arXiv:1001.5443 [hep-ph]];
  E.~Dudas, S.~Lavignac and J.~Parmentier,
  Phys.\ Lett.\ B {\bf 698} (2011) 162
  [arXiv:1011.4001 [hep-th]].


\bibitem{Kitano:2006xg}
  R.~Kitano, H.~Ooguri and Y.~Ookouchi,
  Phys.\ Rev.\ D {\bf 75} (2007) 045022
  [arXiv:hep-ph/0612139];
%
  B.~K.~Zur, L.~Mazzucato and Y.~Oz,
  JHEP {\bf 0810} (2008) 099
  [arXiv:0807.4543 [hep-ph]];
%
  A.~Giveon, A.~Katz and Z.~Komargodski,
  JHEP {\bf 0907} (2009) 099
  [arXiv:0905.3387 [hep-th]];
%
  S.~A.~Abel, J.~Jaeckel and V.~V.~Khoze,
  Phys.\ Lett.\ B {\bf 682} (2010) 441
  [arXiv:0907.0658 [hep-ph]];
  M.~Bertolini, L.~Di Pietro and F.~Porri,
  JHEP {\bf 1201}, 158 (2012)
  [arXiv:1111.2307 [hep-th]].

\bibitem{Azeyanagi:2011uc}
  T.~Azeyanagi, T.~Kobayashi, A.~Ogasahara and K.~Yoshioka,
  JHEP {\bf 1109} (2011) 112
  [arXiv:1106.2956 [hep-ph]].


\bibitem{Komargodski:2009pc}
  Z.~Komargodski and N.~Seiberg,
  JHEP {\bf 0906} (2009) 007
  [arXiv:0904.1159 [hep-th]];
  K.~R.~Dienes and B.~Thomas,
  Phys.\ Rev.\ D {\bf 81} (2010) 065023
  [arXiv:0911.0677 [hep-th]].


\bibitem{Shih:2007av} 
  D.~Shih,
  JHEP {\bf 0802} (2008) 091
  [hep-th/0703196];
  Z.~Sun,
  JHEP {\bf 0901}, 002 (2009)
  [arXiv:0810.0477 [hep-th]];
%
  D.~Curtin, Z.~Komargodski, D.~Shih and Y.~Tsai,
  arXiv:1202.5331 [hep-th].



\bibitem{Fayet:1974jb}
  P.~Fayet and J.~Iliopoulos,
  Phys.\ Lett.\ B {\bf 51} (1974) 461.



\bibitem{Matos:2009xv}
  L.~F.~Matos,
  arXiv:0910.0451 [hep-ph].

\bibitem{Affleck:1983mk}
  I.~Affleck, M.~Dine and N.~Seiberg,
  Nucl.\ Phys.\ B {\bf 241} (1984) 493;
%
  Nucl.\ Phys.\ B {\bf 256} (1985) 557.


\bibitem{Elvang:2009gk}
  H.~Elvang and B.~Wecht,
  JHEP {\bf 0906} (2009) 026
  [arXiv:0904.4431 [hep-ph]];
%
  T.~T.~Dumitrescu, Z.~Komargodski and M.~Sudano,
  JHEP {\bf 1011} (2010) 052
  [arXiv:1007.5352 [hep-th]].


\bibitem{Ferretti:2007ec}
  L.~Ferretti,
  JHEP {\bf 0712} (2007) 064
  [arXiv:0705.1959 [hep-th]];
%
  L.~M.~Carpenter, M.~Dine, G.~Festuccia and J.~D.~Mason,
  Phys.\ Rev.\ D {\bf 79} (2009) 035002
  [arXiv:0805.2944 [hep-ph]].









\end{thebibliography}
\end{document}